\documentstyle[twoside]{article}
\def\runninghead#1#2{\pagestyle{myheadings}
\markboth{\hfill{\protect\footnotesize\it{\quad #1}}}
{{\protect\footnotesize\it{#2\quad}}\hfill}}
\begin{document}

\runninghead{\bf Chubykalo  et al.}
{\bf Helmholtz theorem}

$$$$
{\large {\bf HELMHOLTZ THEOREM AND THE V-GAUGE IN THE PROBLEM OF SUPERLUMINAL AND INSTANTANEOUS SIGNALS IN CLASSICAL ELECTRODYNAMICS}
\bigskip
\bigskip

$\;\;\;\;\;\;\;\;\;\;${\bf A. Chubykalo, A. Espinoza, R. Alvarado Flores$^a$

$\;\;\;\;\;\;\;\;\;\;$and A. Gutierrez Rodriguez}

\bigskip

$\;\;\;\;\;\;\;\;\;\;${\it Escuela de F\'{\i}sica, Cuerpo Academico ``Particulas, campos 

$\;\;\;\;\;\;\;\;\;\;$y astrof\'{\i}sica", Universidad Aut\'onoma
de Zacatecas,

$\;\;\;\;\;\;\;\;\;\;$Apartado Postal C-580\, Zacatecas 98068, ZAC., M\'exico 

$\;\;\;\;\;\;\;\;\;\;$e-mail: achubykalo@yahoo.com.mx}

\bigskip

$\;\;\;\;\;\;\;\;\;\;^a\;${\it Centro de estudios multidisciplinarios, 

$\;\;\;\;\;\;\;\;\;\;$Universidad Aut\'onoma de Zacatecas}

\bigskip

$\;\;\;\;\;\;\;\;\;\;$Received September 24, 2005

\baselineskip 5mm

\bigskip

$$$$
In this work  we substantiate the applying of the Helmholtz
vector decomposition theorem (H-theorem) to  vector fields in
classical electrodynamics. Using the H-theorem, within the
framework of the two-parameter Lorentz-like gauge (so called {\it
v-gauge}), we show that  two kinds of magnetic vector potentials
exist:  one of them (solenoidal) can act exclusively with the
velocity of light $c$ and the other  one (irrotational) with an
arbitrary finite velocity $v$ (including a velocity {\it more}
than $c$) . We show also that the irrotational component of the
electric field has a physical meaning and can propagate
exclusively {\it instantaneously}.


$$$$
{\large {\bf 1. INTRODUCTION}}

\bigskip
\noindent
Lately the use of  a two-parameter Lorentz-like gauge (so called
{\it v-gauge}, see, e.g., [1-5])   in classical electrodynamics
gained popularity among physicists. Most likely one can explain
this by attempting to provide, from the classical electrodynamics
point of view an explanation of superluminal signals detected in a
series of well-known experiments, performed at Cologne \cite{Col},
Berkeley \cite{Ber}, Florence \cite{Flo} and Viena \cite{Vie},
experiments by Tittel {\it et al} \cite{Tit} which revealed that
evanescent waves (in undersized waveguides, e.g.) seem to spread
with a superluminal group velocity. For example, in recent
experiments by Mugnai {\it et al}
 \cite{Mug} superluminal behavior in the propagation of microwaves
 (centimeter wavelenth) over much longer distances (tens of centimeters)
 at a speed 7\% faster than $c$ was reported.

For example, in the recent work \cite{Dmi}   by using the
two-parameter Lorentz-like gauge  ({\it v-gauge} [1-4]) and  using
the Helmholtz theorem it was shown that within the framework of
classical electrodynamics   the instantaneous action at a distance
can exist (scalar potential acts {\it instantaneously} while the
vector  potential propagates at the speed of light) that
implicitly confirms  results of the works [12-14] (in these works
the possibility of the existence of instantaneous action at a
distance was rationalized out of the framework of the {\it
v-gauge} theory).  However the author of \cite{Dmi} does not
substantiate the defensibility of the use of the Helmholtz vector
decomposition theorem for time-dependent vector fields:  the
point is that recently J.  A.  Heras \cite{Her} showed that there
is an inconsistent mathematical procedure here, which is due to
the common misconception that the standard Helmholtz theorem
\cite{Arf} (which allows us to write ${\bf E}= {\bf E}_{i}+{\bf
E}_{s}$,  where ${\bf E}_{i}$ and ${\bf E}_{s}$ are irrotational
and solenoidal components of the vector ${\bf E}$) can be applied
to retarded (time-dependent) vector fields. In other words, when
one introduces the time dependence into a vector field ${\bf E}$
and requires a decomposition of ${\bf E}$ into integral components
one {\it must} prescribe the propagator. For electric and magnetic
fields obeying Maxwell's equations, the causal propagator  is the
retarded Green function $D_R(x-x^{\prime})$, where $x$ is 4
variables $(x,y,z,t)$. Thus, by Heras \cite{Her}, the Helmholtz
theorem for time-dependent vector fields must be formulated (using
Heras' notation with  $c=1$) as follows: A time-dependent
(retarded) vector field ${\bf E}(x)$ vanishing at spatial infinity
is decomposed into {\it three} components {\it irrotational}, {\it
solenoidal} and {\it temporal} one:  ${\bf E}= \tilde{{\bf
E}}_{i}+\tilde{{\bf E}}_{s}+ \tilde{\bf E}_t$, where

\begin{equation}
\tilde{{\bf E}}_{i}(x)=-\nabla\int
D_R(x-x^{\prime})\nabla^{\prime}\cdot {\bf E}(x^{\prime})d^4
x^{\prime},
\end{equation}

\begin{equation}
\tilde{{\bf E}}_{s}(x)=\nabla\times\int
D_R(x-x^{\prime})\nabla^{\prime}\times{\bf E}(x^{\prime})d^4
x^{\prime},
\end{equation}

\begin{equation}
\tilde{\bf E}_t(x)=\frac{\partial}{\partial t}\int
D_R(x-x^{\prime})\frac{\partial{\bf E}(x^{\prime})}{\partial
t^{\prime}}d^4 x^{\prime}.
\end{equation}
In its  standard formulation $({\bf E}={\bf E}_i+{\bf E}_s)$
\cite{Arf}, Heras specifies (see footnote 2 in \cite{Her}) that
Helmholtz theorem can consistently be applied to time-dependent
vector  fields {\it only} (!) when an instantaneous propagation
for the fields is assumed.

Therefore results obtained in \cite{Dmi}  {\it could be} incorrect
if one takes into account the inferences of \cite{Her}.
Nevertheless, taking into account this possible impropriety  in
\cite{Dmi}, we have to note the following:  the inferences of J.
A. Heras \cite{Her} can be incorrect at least in the case of   the
time-dependent electric field written by means of scalar and
vector potentials in the Coulomb gauge. It is obvious that for the
electric field

\begin{equation}
{\bf E}({\bf r},t)=-\nabla\varphi({\bf r},t)
-\frac{1}{c}\frac{\partial{\bf A}({\bf r},t)}{\partial t}
\end{equation}
in this case   an instantaneous propagation  {\it is not} assumed
because the field ${\bf E}$ in (4) can be a retarded solution of
the wave equation

\begin{equation}
\nabla^2{\bf E}-\frac{1}{c^2}\frac{\partial^2{\bf E}}{\partial
t^2}= 4\pi\left(\nabla\varrho +\frac{1}{c^2}\frac{\partial{\bf
j}}{\partial t} \right).
\end{equation}
Accordingly,  it is clear that  here although the electric field
(4)  can be {\it retarded}, it is decomposed into just {\it two}
parts, one of which is {\it pure irrotational} and the other is
{\it pure solenoidal}:

\begin{equation}
{\bf E}={\bf E}_i +{\bf E}_s\;,\;{\rm where} \; {\bf E}_i=
-\nabla\varphi({\bf r},t)\;{\rm and}\; {\bf
E}_s=-\frac{1}{c}\frac{\partial{\bf A}({\bf r},t)}{\partial t},
\end{equation}
(in the Coulomb gauge $\nabla\cdot{\bf A}=0$). This alone shows
that  the inference of J.A. Heras \cite{Her} that a retarded
field cannot be decomposed  into {\it only} two parts
(irrotational and solenoidal) can be insufficiently rigorous. Note also that in his recent work  F. Rohrlich \cite{Roh2} has brought out clearly that the Standard Helmholtz theorem can be applied to time-dependent (retarded) vector fields.
 
$$$$
{\large {\bf 2. TWO KINDS OF MAGNETIC VECTOR POTENTIAL}}

\bigskip
\noindent
In our calculations we use generalized gauge condition (the so
called {\it v-gauge})

\begin{equation}
\nabla\cdot{\bf A}+\frac{c}{v^2}\frac{\partial\varphi}{\partial
t}=0,
\end{equation}
the using of which in classical electrodynamics is already
well-founded (see, e.g.,[1-5]). Here $v$ is some {\it arbitrary}
velocity of propagation for electromagnetic potentials (and it is
not necessarily that $v$ has to be equal to $c$).  In
the Maxwell equations if we express  {\bf E} and {\bf B}  through
potentials,  taking into account {\it v-gauge} (7) and after simple
transformations we obtain:

\begin{equation}
\nabla^2\varphi-\frac{1}{v^2}\frac{\partial^2\varphi}{\partial
t^2} = -4\pi\varrho
\end{equation}
and
\begin{equation}
\nabla^2{\bf A}-\frac{1}{c^2}\frac{\partial^2{\bf A}}{\partial^2
t}=
\left(\frac{v^2-c^2}{cv^2}\right)\nabla\frac{\partial\varphi}{\partial
t}- \frac{4\pi}{c}{\bf j}.
\end{equation}

Let us  start from Eq. (9) which, taking into account Eq. (7), can
be written as

\begin{equation}
\nabla^2{\bf A} - \frac{1}{c^2}\frac{\partial^2{\bf A}}{\partial^2
t}+ \left(\frac{v^2-c^2}{c^2}\right)\nabla(\nabla\cdot{\bf A})=
-\frac{4\pi}{c}{\bf j}
\end{equation}
or, using the identity $\nabla(\nabla\cdot{\bf
A})=\nabla\times(\nabla\times{\bf A})+\nabla^2{\bf A}$, and, multiplying by $\frac{c^2}{v^2}$, we obtain:
\begin{equation}
\nabla^2{\bf A}-\frac{1}{v^2}\frac{\partial^2{\bf A}}{\partial^2
t}+ \left(\frac{v^2-c^2}{v^2}\right)\nabla\times(\nabla\times{\bf
A}) =-\frac{4\pi c}{v^2}{\bf j}.
\end{equation}

Now let the vectors ${\bf A}$ and ${\bf j}$ satisfy the conditions
of the Helmholtz theorem.  So
\begin{equation}
{\bf A}({\bf r},t)={\bf A}_{s}({\bf r},t)+{\bf A}_{i}({\bf
r},t),\quad{\rm and}\quad {\bf j}({\bf r},t)={\bf j}_{s}({\bf
r},t)+{\bf j}_{i}({\bf r},t).
\end{equation}

Now after substituting Eqs.(12) into (10) and (11) we have
respectively:

\begin{eqnarray}
\lefteqn{\nabla^2({\bf A}_{s}+{\bf A}_{i}) -
\frac{1}{c^2}\frac{\partial^2({\bf A}_{s}+{\bf A}_{i})}{\partial^2
t}+ }\nonumber\\ & & +\left(\frac{v^2-c^2}{c^2}\right)\nabla(\nabla\cdot({\bf
A}_{s}+{\bf A}_{i}))=   -\frac{4\pi}{c}({\bf j}_s+{\bf j}_i)
\end{eqnarray}
and

\begin{eqnarray}
\lefteqn{\nabla^2({\bf A}_{s}+{\bf A}_{i})-
\frac{1}{v^2}\frac{\partial^2({\bf A}_{s}+{\bf A}_{i})}{\partial^2
t}+ }\nonumber\\ & & + \left(\frac{v^2-
c^2}{v^2}\right)\nabla\times(\nabla\times({\bf A}_{s}+{\bf
A}_{i}))=  -\frac{4\pi c}{v^2}({\bf j}_s+{\bf j}_i).
\end{eqnarray}

By virtue of the uniqueness of the decomposition of vectors  into
solenoidal and irrotational parts (see \cite{Tik}, e.g.) one can
equate solenoidal components of {\it lhs} and {\it rhs} of Eq.(13)
and irrotational components of {\it lhs} and {\it rhs} of Eq.
(14).   The resultant equations are:

\begin{equation}
\nabla^2{\bf A}_s -\frac{1}{c^2}\frac{\partial^2{\bf
A}_s}{\partial t^2}= -\frac{4\pi}{c}{\bf j}_s
\end {equation}
and
\begin{equation}
\nabla^2{\bf A}_i -\frac{1}{v^2}\frac{\partial^2{\bf
A}_i}{\partial t^2}= -\frac{4\pi c}{v^2}{\bf j}_i.
\end {equation}

Thus one can see that two kinds of  magnetic vector potential
exist: one of which (${\bf A}_s)$ propagates exclusively with the
velocity of light $c$ and the other one with an arbitrary
velocity $v$ (including $v>c$). Note that for $v\rightarrow\infty$
the vector potential ${\bf A}_i$  vanishes within the framework of
the conditions of the Helmholtz theorem  (in accordance with the
assertion of \cite{Dmi}). However one can see that ${\bf A}_i$
exists for $c<v<\infty$.
$$$$

{\large {\bf 3. TWO KINDS OF ELECTRIC FIELD}}

\bigskip
\noindent
Note, {\it however},  the following very important ``feature":  in
the {\it v-gauge} the irrotational part of the electric field (4)
can  propagate instantaneously only!

Indeed if we act by the operator ``$-$grad" to Eq.  (8) we obtain:

\begin{equation}
\nabla^2{\bf E}_\varphi- \frac{1}{v^2}\frac{\partial^2{\bf
E}_\varphi}{\partial t^2} = 4\pi\nabla\varrho,
\end{equation}
where ${\bf E}_\varphi=-\nabla\varphi$  is a field produced
exclusively by means of the electric potential from (4). Next we
rewrite (4) in the following form:

\begin{equation}
{\bf E}({\bf r},t)={\bf E}_\varphi({\bf r},t)
-\frac{1}{c}\frac{\partial{\bf A}_i({\bf r},t)}{\partial t}
-\frac{1}{c}\frac{\partial{\bf A}_s({\bf r},t)}{\partial t}
\end{equation}
or
\begin{equation}
{\bf E}({\bf r},t)= {\bf E}_i({\bf r},t) +{\bf E}_s({\bf r},t),
\end{equation}
where, obviously
\begin{equation}
{\bf E}_i({\bf r},t) ={\bf E}_\varphi({\bf r},t)-
\frac{1}{c}\frac{\partial{\bf A}_i({\bf r},t)}{\partial t}
\end{equation}
and
\begin{equation}
{\bf E}_s({\bf r},t)=-\frac{1}{c}\frac{\partial{\bf A}_s({\bf
r},t)}{\partial t}.
\end{equation}
Let us now act by the operator
``$-\frac{1}{c}\frac{\partial}{\partial t}$" to the Eq. (16):

\begin{equation}
\nabla^2\left\{-\frac{1}{c}\frac{\partial{\bf A}_i({\bf
r},t)}{\partial t}\right\}-\frac{1}{v^2}\frac{\partial^2}{\partial
t^2} \left\{-\frac{1}{c}\frac{\partial{\bf A}_i({\bf
r},t)}{\partial t}\right\}= \frac{4\pi}{v^2}\frac{\partial{\bf
j}_i}{\partial t},
\end{equation}
finally, summing (22) and (17) and taking into account (20) we
obtain

\begin{equation}
\nabla^2{\bf E}_i- \frac{1}{v^2}\frac{\partial^2{\bf
E}_i}{\partial t^2} = 4\pi\left(\nabla\varrho
+\frac{1}{v^2}\frac{\partial{\bf j}_i}{\partial t} \right).
\end {equation}
It is obvious\footnote{\large applying the Helmholtz theorem to the
Maxwell equation $\nabla\times{\bf B}=
\frac{1}{c}\frac{\partial{\bf E}}{\partial t}+\frac{4\pi}{c}{\bf
j}$, after time differentiation we obtain $\frac{\partial^2{\bf
E}_i}{\partial t^2}= -4\pi\frac{\partial{\bf j}_i}{\partial t}$.}
that this expression comes to $ \nabla^2{\bf E}_i=4\pi\nabla\varrho$.

Correspondingly for ${\bf E}_s$ we have:

\begin{equation}
\nabla^2{\bf E}_s- \frac{1}{c^2}\frac{\partial^2{\bf
E}_s}{\partial t^2} = \frac{4\pi}{c^2}\frac{\partial{\bf
j}_s}{\partial t}.
\end{equation}

Thus we see that the vector fields ${\bf E}_i$ and ${\bf E}_s$ are
solutions of the {\it different} equations with ${\bf
E}_i$-``wave" propagating {\it instantaneously} and ${\bf
E}_s$-wave propagating with the velocity $c$ respectively.

By virtue of the uniqueness of the decomposition of vectors into
solenoidal and irrotational parts, the values of ${\bf E}_i$ and ${\bf
E}_s$ cannot depend on a gauge. To verify this let us now
construct the wave equation for the field ${\bf E}$ from Maxwell
equations

\begin{equation}
{\rm rot}\,{\bf B}=\frac{4\pi}{c}{\bf
j}+\frac{1}{c}\frac{\partial{\bf E}}{\partial t},\; {\rm
rot}\,{\bf E}=-\frac{1}{c}\frac{\partial{\bf B}}{\partial t},
\quad{\rm and}\quad {\rm div}\, {\bf E}=4\pi\varrho :
\end{equation}

\begin{equation}
\nabla^2{\bf E}-\frac{1}{c^2}\frac{\partial^2{\bf E}}{\partial
t^2}= 4\pi\left(\nabla\varrho +\frac{1}{c^2}\frac{\partial{\bf
j}}{\partial t} \right).
\end{equation}
Then after applying the Helmholtz theorem to the vectors ${\bf E}$
and ${\bf j}$ in (26) we can equate solenoidal and irrotational
parts of {\it lhs} and {\it rhs} of (26) respectively by virtue of
the uniqueness of the decomposition of vectors in  accordance with
the Helmholtz theorem.  The resultant equations are (see footnote
1):

\begin{equation}
\nabla^2{\bf E}_i- \frac{1}{c^2}\frac{\partial^2{\bf
E}_i}{\partial t^2} = 4\pi\left(\nabla\varrho
+\frac{1}{c^2}\frac{\partial{\bf j}_i}{\partial t} \right)\quad
\Longrightarrow\quad \nabla^2{\bf E}_i=4\pi\nabla\varrho,
\end{equation}

\begin{equation}
\nabla^2{\bf E}_s- \frac{1}{c^2}\frac{\partial^2{\bf
E}_s}{\partial t^2} = \frac{4\pi}{c^2}\frac{\partial{\bf
j}_s}{\partial t}.
\end{equation}

In that way the ``paradox" that ${\bf E}_i$ can be {\it
simultaneously} retarded and istantaneous, observed by Rohrlich
(see Eqs. 3.12-3.17 in \cite{Roh}), is resolved: one can see that
${\bf E}_i$ must be exclusively {\it instantaneous}.

Let us consider the case when exlusively ${\bf E}_i$ can be
responsible for a signal transfer from one point charge $q$ to the
other point charge $Q$ (or to some fixed  point of observation).

Let us suppose the charge $q$ is vibrating by means of some non-electrical
force along the $X$-axis, then  charge $Q$ (or the fixed point
of observation), lying at the same axis at some fixed distance from the charge $q$
vibration centre, will  obviously 
``know" that  charge $q$ is vibrating: in the observation point
the value of the energy density $w$ (which is a point function of
${\bf E}$  at  the $X$-axis) will also oscillate.

Let us now  analyse equations for ${\bf E}_i$, ${\bf E}_s$, ${\bf
j}_i$ and ${\bf j}_s$:

\begin{equation}
\nabla\cdot{\bf E}_i=4\pi\varrho,
\end{equation}
\begin{equation}
\frac{\partial {\bf E}_i}{\partial t}=-4\pi{\bf j}_i,
\end{equation}
and for solenoidal components
\begin{equation}
\nabla\times{\bf E}_s=-\frac {1}{c}\frac{\partial{\bf B}}{\partial
t},
\end{equation}
\begin{equation}
\nabla\times{\bf B}=\frac{1}{c}\frac{\partial{\bf E}_s}{\partial
t}+\frac{4\pi}{c}{\bf j}_s.
\end{equation}
From (31) and (32) we obtain the wave equations:
\begin{equation}
\nabla^2{\bf E}_s- \frac{1}{c^2}\frac{\partial^2{\bf
E}_s}{\partial t^2} = \frac{4\pi}{c^2}\frac{\partial{\bf
j}_s}{\partial t}.
\end{equation}
\begin{equation}
\nabla^2{\bf B}- \frac{1}{c^2}\frac{\partial^2{\bf B}}{\partial
t^2} = -\frac{4\pi}{c}\nabla\times{\bf j}_s.
\end{equation}
Here one can see that the solenoidal components of the
electromagnetic field are in charge of the electromagnetic
radiation with the derivatives of ${\bf j}_s$ as a source of these
waves.
 Let us consider now the field created by a point charge with an arbitrary movement:
\begin{equation}
{\bf r}_q={\bf r}_q(t),
\end{equation}
\begin{equation}
{\bf v}_q={\bf v}_q(t)=\frac{d{\bf r}_q(t)}{dt}.
\end{equation}
The charge density and current density are given as
\begin{equation}
\varrho({\bf r}, t)=q\delta({\bf r}-{\bf r}_q(t)),
\end{equation}
\begin{equation}
{\bf j}({\bf r},t)=q{\bf v}_q\delta({\bf r}-{\bf r}_q(t)).
\end{equation}
These quantities are not independent:
\begin{equation}
\frac{\partial\varrho}{\partial t}+\nabla\cdot{\bf j}=0.
\end{equation}

Let us find the irrotational and solenoidal components of the
current density. From the Helmholtz theorem we have:
\begin{equation}
{\bf j}_i({\bf
r},t)=-\frac{1}{4\pi}\nabla\int\frac{\nabla^{\prime}\cdot{\bf
j}({\bf r}^{\prime},t)} {|{\bf r}-{\bf r}^{\prime}|}dV^{\prime}.
\end{equation}
Taking into account Eq. (39) we obtain
\begin{equation}
{\bf j}_i({\bf r},t)=\frac{1}{4\pi}\nabla\frac{\partial}{\partial
t}\int\frac{\varrho({\bf r}^{\prime},t)}{|{\bf r}-{\bf
r}^{\prime}|}dV^{\prime},
\end{equation}
Finally, substituting (37) we have:
\begin{equation}
{\bf j}_i({\bf r},t)=\frac{q}{4\pi}\nabla\frac{\partial}{\partial
t}\frac{1}{|{\bf r}-{\bf r}_q(t)|}.
\end{equation}
 One can rewrite this expression in the following form
\begin{equation}
{\bf j}_i({\bf r},t)=-\frac{1}{4\pi}\frac{3{\bf n}(\tilde{\bf
j}\cdot{\bf n})-\tilde{\bf j}} {|{\bf r}-{\bf r}_q(t)|^3},
\end{equation}
where
\begin{equation}
\tilde{\bf j}= q{\bf v}_q(t),
\end{equation}
\begin{equation}
{\bf n}=\frac{{\bf r}-{\bf r}_q(t)}{|{\bf r}-{\bf r}_q(t)|}.
\end{equation}
Pay attention to the similitude of Eq. (43) and the well-known
expression for the electric field created by a dipole.

On the other hand for the solenoidal component we have:
\begin{equation}
{\bf j}_s({\bf
r},t)=\frac{1}{4\pi}\nabla\times\int\frac{\nabla^{\prime}\times{\bf
j}({\bf r}^{\prime},t)}{|{\bf r}-{\bf r}^{\prime}|}dV^{\prime}.
\end{equation}
After some calculations we obtain:
\begin{equation}
{\bf j}_s({\bf r},t)=q{\bf v}_q\delta({\bf r}-{\bf r}_q(t))+
\frac{1}{4\pi}\frac{3{\bf n}(\tilde{\bf j}\cdot{\bf n})-\tilde{\bf
j}} {|{\bf r}-{\bf r}_q(t)|^3}.
\end{equation}
So, comparing (47) and (43), we conclude that
\begin{equation}
{\bf j}_s({\bf r},t)=-{\bf j}_i({\bf r},t)
\end{equation}
at every point with the exception of the point of location of the
charge.

The obtained expressions permit us to  find the irrotational
component ${\bf E}_i$ of the electric field created by the charge.
Comparing (30) and (42) we obtain
\begin{equation}
{\bf E}_i=q\frac{{\bf r}-{\bf r}_q(t)}{|{\bf r}-{\bf r}_q(t)|^3}.
\end{equation}
One can see that the field ${\bf E}_i$ is a Coulomb type field: it
is conservative and has a spherical symmetry with respect to the
instantaneous location of the charge. Besides the field ${\bf
E}_i$ ``moves" (changes) instantaneously everywhere in space
together with the charge.

As an example consider the case when the point charge $q$ performs
harmonic oscillations along the $X$-axis:
\begin{equation}
{\bf r}_q(t)=(A_0\sin \omega t){\bf i},
\end{equation}
where ${\bf i}$ is the unit vector in the positive direcction of
the $X$-axis. Using (42) and (48) we obtain  for any
point\footnote{\large Except for the location point of $q$, of course.}
 at the $X$-axis (${\bf r}=x{\bf i}$):
\begin{equation}
{\bf j}_i=-{\bf j}_s=-\frac{q}{4\pi}\frac{A_0\omega\cos\omega
t}{|x-A_0\sin\omega t|}{\bf i}
\end{equation}
and
\begin{equation}
{\bf E}_i=q\frac{x- A_0\sin\omega t}{|x- A_0\sin\omega t|^3}{\bf
i}.
\end{equation}
So one can see that ${\bf E}_i$ is directed along  the $X$-axis
at the $X$-axis.

In order to determine the solenoidal component  ${\bf E}_s$ we do
the following. The field ${\bf E}$ created by the charge has to be
periodic, consequently, we can develop its solenoidal and
irrotational components  in the Fourier series:
\begin{equation}
{\bf E}_i({\bf r},t)=\sum_{n=0}^\infty{\bf E}_{in}e^{-in({\bf
k}\cdot{\bf r}-\omega t)},
\end{equation}
\begin{equation}
{\bf E}_s({\bf r},t)=\sum_{n=0}^\infty{\bf E}_{sn}e^{-in({\bf
k}\cdot{\bf r}-\omega t)}.
\end{equation}
From the properties of these fields
\begin{equation}
\nabla\times{\bf E}_i=0,
\end{equation}
\begin{equation}
\nabla\cdot{\bf E}_s=0
\end{equation}
we obtain for every $n$
\begin{equation}
{\bf k}\times{\bf E}_{in}=0,
\end{equation}
\begin{equation}
{\bf k}\cdot{\bf E}_{sn}=0.
\end{equation}
The last equations mean that the vectors ${\bf E}_i$ and ${\bf E}_s$ must
be mutually perpendicular everywhere in space and thus  ${\bf
E}_s$ must be perpendicular to the $X$-axis in every point of the
$X$-axis (${\bf E}_i$ (Eq. (52) is collinear to $X$-axis).

On account of the symmetry of the problem and because of ${\bf
E}={\bf E}_i+{\bf E}_s$,  ${\bf E}_s$ must be equal to {\it zero}
along of the $X$-axis. It can mean solely the following: The
irrotational component of the electric field has a physical
meaning and in some cases is charged with the instantaneous energy
and momentum transmission.

So we have made sure that the irrotational component of the
electric field has a physical meaning and in some cases,
obviously, {\it  it} is solely  in charge of the energy and
momentum transmissions which, evidently, have to be instantaneous
in this case (see Eqs. (23), (27) and footnote 1). It is obvious also that this
field cannot be directly obtained from the well-known expression
for the electric field created by an arbitrarily moving charge

\begin{eqnarray}
   \lefteqn{{\bf E}({\bf r},t)=q\left\{\frac{({\bf R}-R\frac{{\bf
         V}}{c})(1-\frac{V^{2}}{c^{2}})}{(R-{\bf R}\frac{{\bf
   V}}{c})^{3}}\right\}_{t_0}+}\nonumber\\
& & +q\left\{\frac{[{\bf R}\times[({\bf
   R}-R\frac{{\bf V}}{c})\times\frac{{\bf{\dot{V}}}}{c^{2}}]]}{(R-{\bf
   R}\frac{{\bf V}}{c})^{3}}\right\}_{t_0},
   \end{eqnarray}
where ${\bf V}={\bf V}(t_0)$ is the velocity of the charge $q$,
$t_0=t-R/c$, $R=|{\bf r}-{\bf r}_q(t_0)|$, because the full field
${\bf E}$ in (59) was obtained from the Li\'enard-Wiechert
potentials  taking into account the retardation, and our field
${\bf E}={\bf E}_i$ at the $X$-axis {\it is not retarded} in
accordance with Eqs. (23), (27). So at the $X$-axis, taking into
account the {\it instantaneousness} of ${\bf E}_i$ we must put a
velocity of the propagation of the field ${\bf E}$ $c=\infty$ and
then we obtain Eq. (52).

$$$$

{\large {\bf 4. CONCLUSION}}

\bigskip
\noindent
One can see that the irrotational part ${\bf
A}_i$ of the vector potential ${\bf A}$  and the scalar
potential $\varphi$ {\it can} propagate with an arbitrary {\it
finite} velocity including a velocity {\it more} than $c$ as well
as instantaneously in the case of the scalar potential (we showed that for $v\rightarrow\infty$
the vector potential ${\bf A}_i$  vanishes within the framework of
the conditions of the Helmholtz theorem).

In regards to the {\it irrotational} conponent of the electric field (see
Eqs. (23) and (27)), it has a physical meaning and can propagate
exclusively {\it instantaneously}. Therefore we can conclude that
there are {\it two} mechanisms of the energy and momentum
transmission in classical electrodynamics:

1) the {\it retarded} one by means of a radiation (${\bf E}_s$ and
${\bf B}$), see Eqs. (33), (34);

2) the {\it instantaneous} one by means of the irrotational field
${\bf E}_i$.

Note that for the describing of an energy transfer in the second
mechanism along  the line of the interaction of two point
charges the use of the Poynting vector concept makes no sense at
all. Note also that ${\bf E}_i$ cannot have any functional
relations with the magnetic field (see Eqs. (29)-(34)). Thus we
see that  field ${\bf E}_i$ although is materially existent,
cannot participate in the phenomenon known as {\it electromagnetic
wave}.

In view of the obvious spherical symmetry and the non-retardation
of the field ${\bf E}_i$ (49) we would like to make a quotation of
P.A.M. Dirac: ``{\sl As long as we are dealing only with
transverse waves, we cannot bring in the Coulomb interactions
between particles. To bring them in, we have to introduce
longitudinal electromagnetic waves...  We thus get a new version
of the theory, in which the electron is always accompanied by the
Coulomb field around it. Whenever an electron is emitted, the
Coulomb field around it is simultaneously emitted, forming a kind
of dressing for the electron. Similarly, when an electron is
absorbed, the Coulomb field around it is simultaneosly absorbed.
This is, of course, very sensible physically, but it also means a
rather big departure from relativistic ideas. For, if you have a
moving electron, then the Coulomb field around it is not
spherically symmetrical\footnote{\large namely the solenoidal field ${\bf
E}_s$ has to be in charge of the ``flatting" of the electric
field of a moving point charge.}. Yet it is the spherically
symmetric Coulomb field that has to be emitted here together with
the electron.}" \cite{Dir}. So there is good reason to believe
that exactly the field ${\bf E}_i$ can play a role of the
 spherically symmetric electric field, which is mentioned by Dirac, which
always accompanies any point charge and it is not a generally
accepted Coulomb field because it depends on time. We would like to
name this field ${\bf E}_i$ ``Dirac's field".

Everything described above  can provide a theoretical rationale (within
the framework of classical electrodynamics)  of a series
of  well-known experiments [6-11] mentioned in the Introduction.
Nevertheless it is significant that one can find a theoretical
rationale of the existence of the superluminal interraction out of
the  framework of the Helmholtz theorem and {\it v-gauge}-theory
in the review works \cite{Rec} and \cite{Kot} (see also the review
\cite{Dvo}).

Finally, we can affirm that applying the
Helmholtz theorem to classical ekectrodynamics allows us to
conclude that in classical electrodynamics so called {\it
instantaneous action at a distance} with the {\it infinite}
velocity of interaction  {\it can take place}  as well as (within
the framework of the {\it v-gauge}-theory) the superluminal action
with a {\it finite} velocity of interaction.

\bigskip
$$$$
{\large {\bf ACKNOWLEDGMENTS}}

\bigskip
\noindent
We are grateful to V. Onoochin  and Dr. V. Hnizdo
for many stimulating discussions and critical comments.

$$$$ {\large {\bf REFERENCES}}

\end{document}